\documentclass[doublecol,linenumbers]{epl2} % for 2 columns style with line numbers
\usepackage[T2A]{fontenc}
\usepackage[english]{babel}
\usepackage{amssymb,latexsym,amsmath,amscd}
\usepackage{graphicx,color,framed}
\title{A flexible polymer chain in a critical solvent: Coil or globule?}
\author{Yu.~�. Budkov \inst{1,2,3} \thanks{E-mail: \email{urabudkov@rambler.ru}}, A. L. Kolesnikov\inst{4,5} \thanks{E-mail: \email{bancocker@mail.ru}}, N. Georgi\inst{6} and M. G. Kiselev\inst{1,3}. }
\shortauthor{Yu.~�. Budkov \etal}
\institute{
    \inst{1} G.A. Krestov Institute of Solution Chemistry of the Russian Academy of Sciences - Ivanovo, Russia\\
    \inst{2} National Research University Higher School of Economics - Moscow, Russia\\
    \inst{3} Lomonosov Moscow State University, Department of Chemistry - Moscow, Russia\\
    \inst{4} Ivanovo State University - Ivanovo, Russia\\
    \inst{5} Institut f\"{u}r Nichtklassische Chemie e.V., Universitat Leipzig - Leipzig, Germany\\
    \inst{6} Max Planck Institute for Mathematics in the Sciences - Leipzig, Germany.
}

\abstract{
We study the behavior of a flexible polymer chain in the presence of a
low-molecular weight solvent in the vicinity of a liquid-gas critical point
within the framework of a self-consistent field theory. The total free energy of the
dilute polymer solution is expressed as a function of the radius of
gyration of the polymer and the average solvent number density
within the gyration volume at the level of the mean-field approximation.
Varying the strength of attraction between polymer and solvent we show that
two qualitatively different regimes occur at the liquid-gas critical point.
In case of weak polymer-solvent interactions the polymer chain is in a globular state.
On the contrary, in case of strong polymer-solvent interactions the polymer chain attains an
expanded conformation. We discuss the influence of the critical solvent density fluctuations
on the polymer conformation. The reported effect could be used to excert control on the polymer
conformation by changing the thermodynamic state of the solvent. It could also be helpful to
estimate the solvent density within the gyration volume of the polymer for drug delivery
and molecular imprinting applications.}
\begin{document}

\maketitle
\section{Introduction}
As has been shown by Koga et al \cite{Koga}, in the vicinity of a liquid-gas
critical point of the solvent a dramatic expansion of a dissolved polymer chain takes place.
Such anomalous conformational behavior of the polymer chain occurs due to strong solvent density fluctuations
which arise in the neighborhood of the liquid-gas critical point. Up to now, only a few theoretical
results have been reported on a thermodynamically stable conformation of a single polymer chain
in a supercritical solvent in the neighborhood of the critical point. Dua and Cherayil \cite{Dua}
have developed a first principles statistical theory of the isolated polymer chain in supercritical solvent
and showed that on approaching the liquid-vapor  critical point, a polymer chain first
collapses (when correlation radius order of a polymer size) and then returns to its initial dimensions
(when correlation radius is much bigger the size of polymer chain). However, the authors fully ignored 
an effect of the solvent density renormalisation near the polymer backbone, although the latter should be 
important in the case of the strong polymer-solvent interactions \cite{Chandler_2002,Budkov1,Budkov2}.
Moreover, authors did not specify the dependence of correlation length on thermodynamic parameters 
of the solvent.

Simmons and Sanchez \cite{Sanchez} published a scaled particle theory for the
coil-globule transition of a chain of attractive hard spheres in an attractive hard sphere solvent.
As was shown by the authors, in the vicinity of the solvent critical point only the collapse
of the polymer chain takes place but not its expansion. In the framework of theory of Simmons and Sanchez
the effect of solvent density renormalization near the polymer chain was taken into account but
the effect of critical solvent density fluctuations was missing. The same result was obtained
by Erukhimovich \cite{Erukhimovich} in the framework of the field-theoretical approach at the 
level of Ginzburg-Landau theory within model of a compressible lattice gas. 
Sumi and co-authors \cite{Sekin} employing a classical density functional theory
showed that on approaching the liquid-vapor critical point along the critical isochore,
the polymer chain can undergo collapse at a so-called crossover temperature which is
slightly higher than critical temperature. Subsequently, below the crossover temperature
near the critical point a very dramatic expansion of the polymer chain takes place.
It was concluded that the dramatic expansion of the polymer chain is related
to the so-called solvent mediated interactions arising from strong solvent
density fluctuations \cite{Fisher_1978}.

Despite the evident success in rationalising the conformational behavior of the polymer
chain in the critical solvent, clear understanding of the different conformational regimes has still not been reached.
How does the  globular or coiled state conformation of the polymer depend on the  microscopic polymer-solvent
interaction parameters in a critical solvent? Addressing this question we develop a simple analytical
self-consistent field theory of an isolated polymer chain immersed in a low-molecular weight solvent.

The presented theory simultaneously takes into account two effects:
\begin{itemize}
\item{effect of solvent density renormalisation near the polymer chain due to polymer-solvent interactions;}
\item{indirect solvent mediated monomer-monomer attractive interaction due to solvent density fluctuations near the critical point
(so-called quasi-Casimir forces).}
\end{itemize}

The  presented study is based on the formalism, developed in our previous works
\cite{Budkov1,Budkov2}. We investigate the conformational behavior of the polymer chain
approaching  the critical point of the solvent along the critical isochore. In the case of a
weak polymer-solvent attraction at the liquid-gas critical point the polymer chain is in
a globular state. When the polymer-solvent attraction  exceeds a threshold value the polymer
undergoes a dramatic expansion at the critical point.

\section{Theory}
We consider an isolated polymer chain immersed in a
low-molecular weight solvent at a specified number
density $\rho$ and temperature $T$. As in our previous
works \cite{Budkov1,Budkov2} we assume for convenience
that the volume of the system consists of two parts: the gyration volume
containing predominantly monomers of the polymer chain and the bulk solution.
Our aim is to study the conformational behavior of a polymer chain in the vicinity 
of the solvent liquid-gas critical point as a function of the polymer-solvent interaction strength.
We also assume that pair potentials of interactions monomer-monomer, monomer-solvent and solvent-solvent 
have a following form
\begin{equation}
\label{eq:pot}
V_{ij}(\bold{r})=\Biggl\{
\begin{aligned}
-\epsilon_{ij}\left(\frac{\sigma_{ij}}{r}\right)^6, \quad& |\bold{r}|> \sigma_{ij}\,\\
\infty,\quad&|\bold{r}|\leq \sigma_{ij},
\end{aligned}
\end{equation}
where $i,j=m,s$; $\sigma_{ij}$ and $\epsilon_{ij}$ are effective diameters and energetic parameters, 
respectively. We consider the theory at the level of the mean-field approximation, however,
taking into account the solvent density fluctuation effect on the monomer-monomer interaction.
We describe the thermodynamics of the bulk solution by the Van-der-Waals equation of state
and use a Flory-type expression for the ideal part of the polymer free energy \cite{Flory_book,Giacometti}.

%We shall use the Van-der-Waals equation of state for the description of the solvent's
%thermodynamics in the bulk solution and Flory-type expression
%for a polymer free energy \cite{Flory_book}.
We would also like to stress that in contrast to our previous works \cite{Budkov1,Budkov2} within 
the present theory we do not introduce the second virial coefficients as parameters of interactions, 
but we construct the total free energy by using different expressions which are straightforwardly 
related to repulsive and attractive parts of interaction potentials.

In order to solve of the posed problem, we start from an appropriate thermodynamic potential of 
a dilute polymer solution which can be expressed in the following form
\begin{equation}
\label{eq:Phi}
\Omega(R_{g},N_{s})=F_{p}(R_{g})+F_{s}(N_{s},R_{g})+PV_{g}-\mu N_{s},
\end{equation}
where $R_{g}$ is a radius of gyration of the polymer chain, $N_{s}$ is the number
of solvent molecules within the gyration volume $V_{g}=4\pi R_{g}^3/3$,
$\mu$ is the chemical potential of the solvent, and $P$ is the pressure of the bulk solution.
The polymer free energy in the framework of the mean-field approximation takes the form
\begin{eqnarray}
\label{eq:Fp}
F_{p}(R_{g})=k_{B}T\left(\frac{9}{4}\left(\alpha^2+\frac{1}{\alpha^2}\right)-\frac{3}{2}\ln{\alpha^2}\right)-\nonumber\\
Nk_{B}T\ln\left(1-\frac{Nv_{m}}{V_{g}}\right)-\frac{a_{p}N^2}{V_{g}},
\end{eqnarray}
where $v_{m}=2\pi\sigma_{m}^3/3$ is the Van-der-Waals volume of monomers 
($\sigma_{m}=\sigma_{mm}$ denotes the monomer's effective diameter),
\begin{equation}
a_{p}=\frac{1}{2}\epsilon_{m}\int\limits_{r>\sigma_{m}}d\bold{r}\left(\frac{\sigma_{m}}{r}\right)^6=v_{m}\epsilon_{m}
\end{equation}
is the Van-der-Waals attraction parameter of monomers, $\epsilon_{m}=\epsilon_{mm}$ is an energetic parameter of the monomer-monomer attraction,
$\alpha=R_{g}/R_{0g}$ denotes the expansion factor, $R_{0g}^2=Nb^2/6$ is the mean-square radius of gyration of the ideal polymer
chain, $N$ is the degree of polymerization and $b$ is the Kuhn length of the segment. The first term in (\ref{eq:Fp}) 
is the free energy of the ideal Gaussian polymer chain within the Fixman approximation \cite{Fixman,Birshtein_1}.
Thus to take into account the repulsive and attractive interactions between monomers 
we use a concept of $"$separate monomers$"$ \cite{Lifshitz1} and introduce the Van-der-Waals 
type expression for excess free energy of the polymer chain. The solvent free energy $F_{s}$ within 
a mean-field approximation can be expressed in the following form
\begin{eqnarray}
\label{eq:Fs}
F_{s}(R_{g},N_{s})=N_{s}k_{B}T\left(\ln{\frac{N_{s}\Lambda_{s}^3}{V_{g}-N_{s}v_{s}-Nv_{ms}}}-1\right)-\nonumber\\
\frac{N_{s}^2a_{s}}{V_{g}}-\frac{NN_{s}a_{ps}}{V_{g}}-
\frac{a_{ps}^2N_{s}^2N^2\chi_{T}}{2V_{g}^3},
\end{eqnarray}
where $v_{s}=2\pi\sigma_{s}^3/3$ is a Van-der-Waals volume of the solvent molecules
($\sigma_{s}=\sigma_{ss}$ is an effective diameter of the solvent molecules), 
$v_{ms}=4\pi \sigma_{ms}^3/3$ ($\sigma_{ms}=(\sigma_{m}+\sigma_{s})/2$)
is an effective volume of monomer and solvent molecule at contact,
\begin{equation}
a_{s}=\frac{1}{2}\epsilon_{s}\int\limits_{r>\sigma_{m}}d\bold{r}\left(\frac{\sigma_{s}}{r}\right)^6=v_{s}\epsilon_{s}
\end{equation}
is a solvent-solvent attraction parameter, $\epsilon_{s}=\epsilon_{ss}$ is an energetic parameter of the solvent-solvent attraction
\begin{equation}
a_{ps}=\epsilon_{ms}\int\limits_{r>\sigma_{ms}}d\bold{r}\left(\frac{\sigma_{ms}}{r}\right)^6=v_{ms}\epsilon_{ms}
\end{equation}
is a monomer-solvent attraction parameter, $\epsilon_{ms}$ is an energetic parameter of the monomer-solvent attraction.
The first term in (\ref{eq:Fs}) describes the ideal contribution and contributions of the solvent-solvent 
and monomer-solvent exluded volume interactions. It should be noted that in (\ref{eq:Fs}) we use the simplest 
possible method to account for the so-called depletion forces \cite{Hansen} which may lead to an additional 
repulsion of the solvent molecules from the gyration volume due to the presence of monomers. Second and third terms correspond
to the solvent-solvent and monomer-solvent attraction, respectively. 
The forth term is a first correction to the mean-field approximation in the 
framework of the cumulant expansion \cite{Budkov1} and describes the contribution 
of solvent mediated monomer-monomer interaction that is due to solvent density fluctuations 
(quasi-Casimir forces); $\chi_{T}$ is the isothermal compressibility of the solvent 
in the gyration volume. The fluctuation correction to the mean-field approximation 
is not related to the third virial contribution, though it has a similar functional form. 
It should be emphasized, that within our theory we deal with two independent order parameters -- 
the radius of gyration of the polymer chain and number of the solvent molecules that are in gyration volume.

The solvent chemical potential $\mu$ and pressure in the bulk solution $P$ in our 
model are determined by the following mean-field expressions
\begin{eqnarray}
\label{eq:mu}
\frac{\mu(\rho,T)}{k_{B}T}=\ln(\rho\Lambda_{s}^3)+\frac{\rho v_{s}}{1-\rho v_{s}}-\ln\left(1-\rho v_{s}\right)-\frac{2 a_{s}\rho}{k_{B}T},
\end{eqnarray}
and
\begin{equation}
\label{eq:Press}
P(\rho,T)=\frac{k_{B}T\rho}{1-\rho v_{s}}-a_{s}\rho^2,
\end{equation}
where $\rho$ is a number density of the solvent in the bulk solution.
The expression (\ref{eq:Press}) is a well known Van-der-Waals equation of state.
We assume that gyration volume is sufficiently large, so that surface layer does not 
contribute into free energy of the solution.

The equilibrium values $N_{s}$ and $R_{g}$ are determined from the minimum conditions
of the thermodynamic potential $\Omega$, i.e. from the equations
\begin{equation}
\label{eq:constr1}
\frac{\partial{\Omega}}{\partial{N}_{s}}=0, ~~ \frac{\partial{\Omega}}{\partial{R_{g}}}=0.
\end{equation}
Substituting (\ref{eq:Phi}) into the equations (\ref{eq:constr1}) and using the expressions
(\ref{eq:Fp}) and (\ref{eq:Fs}) we arrive at the following system of coupled equations
\begin{eqnarray}
\label{eq:alpha}
&\alpha^5-\frac{2}{3}\alpha^3-\alpha =
\frac{3\sqrt{6}}{\pi b^3}\sqrt{N}\left(\frac{v_{m}}{1-\frac{9\sqrt{6}v_{m}}{2\pi\sqrt{N}\alpha^3b^3}}-\beta a_{p}-B\right)-\nonumber\\
&\frac{2}{3}N\beta a_{ps}\rho_{s}\alpha^3-\frac{2\pi N^{3/2}\beta\alpha^{6}b^3}{81}\left(P-P_{g}\right),
\end{eqnarray}
and
\begin{eqnarray}
\label{eq:rho}
\rho_{s}=\rho e^{\frac{9\sqrt{6} a_{ps}}{2\pi\sqrt{N}\alpha^3b^3 k_{B}T}-
\frac{\mu_{ex,g}(\rho_{s},\alpha,T)-\mu_{ex,b}(\rho,T)}{k_{B}T}},
\end{eqnarray}
where we have introduced the solvent number density in the gyration volume $\rho_{s}=N_{s}/V_{g}$,
the excess chemical potentials of the solvent $\mu_{ex,g}$, $\mu_{ex,b}$
in the gyration volume and in the bulk solution, respectively, and the inverse
temperature $\beta =1/k_{B}T$. The solvent pressure in the gyration volume $P_{g}$
is given by the following expression
\begin{equation}
\label{eq:PressGyr}
P_{g}=\frac{\rho_{s}k_{B}T}{\gamma-\rho_{s}v_{s}}-a_{s}\rho_{s}^2,
\end{equation}
where $\gamma=1-\frac{9\sqrt{6}v_{ms}}{2\pi\sqrt{N}\alpha^3b^3}$.
The value $B$ determines the influence of the solvent density fluctuations on
the value of radius of gyration and has the following form
\begin{eqnarray}
B=\frac{a_{ps}^2\rho_{s}}{\left(k_{B}T\right)^2}\frac{(\gamma-\rho_{s}v_{s})\left(\gamma(\gamma-2\rho_{s}v_{s})-
\frac{\rho_{s}a_{s}}{k_{B}T}(\gamma-\rho_{s}v_{s})^3\right)}{\left(\gamma-\frac{2a_{s}\rho_{s}}{k_{B}T}(\gamma-\rho_{s}v_{s})^2\right)^2}.
\end{eqnarray}

The excess chemical potential of the solvent in the gyration volume can be obtained by the following relation
\begin{eqnarray}
\label{eq:muGyr}
\mu_{ex,g}(\rho_{s},\alpha,T)=k_{B}T\left(\frac{\rho_{s} v_{s}}{\gamma-\rho_{s} v_{s}}-
\ln\left(\gamma-\rho_{s} v_{s}\right)\right)-\nonumber\\
2 a_{s}\rho_{s}-\frac{243 a_{ps}^2\gamma}{4\pi^2 N\alpha^6 b^6k_{B}T}\frac{(\gamma-\rho_{s}v_{s})(\gamma-3\rho_{s}v_{s})}{\left(\gamma-\frac{2a_{s}\rho_{s}}{k_{B}T}(\gamma-\rho_{s}v_{s})^2\right)^2}.
\end{eqnarray}
The first and second terms in the expression (\ref{eq:muGyr}) determine the mean-field contribution to the excess chemical
potential of the exluded volume and the attraction interactions, respectively. The third term determines the contribution 
of the solvent density fluctuations in the gyration volume.

It should be noted that similar pure mean-field theories have been developed by Erukhimovich \cite{Erukhimovich} 
for the case of infinitely long polymer chain and by Simmons and Sanchez \cite{Sanchez} for the polymer chain 
of finite degree of polymerization. However, in both works only collapse of the polymer chain in the vicinity of the 
liquid-vapor critical point has been discussed. In order to describe the coil-globule transition in the critical 
solvent Erukhimovich used a model of a compressible lattice gas at the level of the mean-field approximation, 
whereas Simmons and Sanchez used a scaled particle theory. However, it is unclear how the phenomenological parameters 
of interaction within the  lattice gas model are related to the  pair potentials of interactions 
between components of the solution. Moreover, to determine a parameter of the polymer-solvent attraction 
$\epsilon_{ms}$ in both works  a standard Berthelot rule $\epsilon_{ms}=\sqrt{\epsilon_{m}\epsilon_{s}}$ 
was used, which obviously cannot be justified for the case of polymer solution. In the present study in contrast 
to work \cite{Erukhimovich} as well as in work \cite{Sanchez} we directly introduce the pair potentials 
 interactions between components of the solution (\ref{eq:pot}) using parameters of interactions 
$\epsilon_{ms}$, $\epsilon_{m}$, and $\epsilon_{s}$ as independent variables. In addition, in contrast 
to works \cite{Erukhimovich,Sanchez} we have taken into account the solvent mediated fluctuation interactions 
between the monomers and obtain a new regime of conformational behavior of the polymer, namely its strong 
expansion when solvent reaches the liquid-vapour critical point (see the next section).

\section{Numerical results and discussions}
For convenience within the numerical calculations we introduce dimensionless densities $\tilde{\rho}_{s}=\rho_{s}v_{s}$
and $\tilde{\rho}=\rho v_{s}$, and dimensionless interaction parameters $s=\sigma_{m}/\sigma_{s}$,
$t=\epsilon_{m}/\epsilon_{s}$, $u=\epsilon_{ms}/\epsilon_{s}$.
Using these definitions we rewrite the equations (\ref{eq:alpha}-\ref{eq:rho}) in the following form
\begin{eqnarray}
\label{eq:alpha2}
&\alpha^5-\frac{2}{3}\alpha^3-\alpha=\frac{3\sqrt{6}\sqrt{N}}{\pi \tilde{b}^3}\left(\frac{s^3}{1-\frac{9\sqrt{6}s^3}{2\pi\sqrt{N}\alpha^3\tilde{b}^3}}-
\frac{ts^3}{\tilde{T}}-\tilde{B}\right)\nonumber\\
&-\frac{\tilde{\rho}_{s}u(1+s)^3N\alpha^3}{6\tilde{b}^3}
-\frac{2\pi N^{3/2}\tilde{b}^3\alpha^{6}}{81\tilde{T}}\left(\tilde{P}-\tilde{P}_{g}\right),
\end{eqnarray}
\begin{equation}
\label{eq:rho2}
\tilde{\rho}_{s}=\tilde{\rho}e^{\frac{9\sqrt{6}u(1+s)^3}{8\pi\sqrt{N}\alpha^3\tilde{b}^3\tilde{T}}+\frac{\tilde{\mu}_{ex,b}(\tilde{\rho},\tilde{T})-\tilde{\mu}_{ex,g}(\tilde{\rho}_{s},\alpha,\tilde{T})}{\tilde{T}}},
\end{equation}
where $\tilde{T}=k_{B}T/\epsilon_{s}$ is a dimensionless temperature, $\tilde{\mu}_{ex,b}=\mu_{ex,b}/\epsilon_{s}$ and $\tilde{\mu}_{ex,g}=\mu_{ex,g}/\epsilon_{s}$
are dimensionless excess chemical potentials of the solvent in the bulk solution and in the gyration volume, respectively; $\tilde{B}=B/v_{s}$;
$\tilde{b}=b/v_{s}^{1/3}$ is a dimensionless Kuhn length of the segment. Moreover,
we have introduced the dimensionless pressures $\tilde{P}=Pv_{s}/\epsilon_{s}$
and $\tilde{P}_{g}=P_{g}v_{s}/\epsilon_{s}$.

Turning to the numerical analysis of the system of equations (\ref{eq:alpha2}--\ref{eq:rho2}) we fix
the index of polymerization $N=10^{3}$, dimensionless Kuhn length of the segment
$\tilde{b}=s\left(\frac{3}{2\pi}\right)^{1/3}$ (so that $b=\sigma_{m}$), and $t=1$.

We first discuss the conformational behavior of the polymer chain approaching
the liquid-gas critical point of the solvent along the critical isochore for different
values of the polymer-solvent attraction parameter $u$. In Fig. 1
two different regimes of the behavior of expansion factor $\alpha$ are shown.
At small values of $u$ (weak polymer-solvent attraction)
the polymer chain collapses at the critical point. However, when the
polymer-solvent attraction is strong a qualitatively different behavior of expansion factor takes place.
Namely, decreasing the temperature along the critical isochore the polymer chain first collapses
at a temperature higher than critical temperature and then subsequently strongly expands
in a small vicinity of the critical point. Such conformational behavior is in agreement with numerical
results of Sumi et al \cite{Sekin} and with experimental results of Koga et al \cite{Koga}.
Fig. 2 shows the expansion factor $\alpha_{c}=\alpha(\tilde{\rho}_{c},\tilde{T}_{c})$
at the critical point as a function of $u$ for different $s$.
In the case of a weak polymer-solvent attraction (small $u$), the expansion factor $\alpha$ in the vicinity of
the liquid-gas critical point is effectively constant and does not depend on $u$.
However, when the parameter $u$ exceeds a certain threshold value the expansion
factor monotonically increases. Such nontrivial behavior of the expansion factor
is related to the fact that at weak polymer-solvent interaction
(small $u$) polymer is "solvophobic". In this case the local number density of
solvent near the polymer chain is much less than the number density in the bulk solution
($\tilde{\rho}_{s}\ll \tilde{\rho}$), so that the polymer collapses due to the pressure difference
between the gyration volume and the bulk solution.

In the case of a sufficiently strong polymer-solvent attractive interaction
the fluctuation contibution to the solvent excess chemical potential becomes important.
The latter leads to an additional effective attraction between polymer backbone and
solvent molecules thus increasing the number density of solvent
within the gyration volume and eventually equilibrates the pressures between
the gyration volume and the bulk solution, so that the polymer chain
becomes "solvophilic". In this case the density of the solvent
within the gyration volume becomes comparable to the solvent density in the
bulk solution ($\tilde{\rho}_{s}\sim \tilde{\rho})$, while polymer chain expands
to a coiled conformation. Further increasing of the polymer-solvent attraction leads to
enhancement the solvent density in gyration volume and shrinking of the polymer coil. 
The latter is related to the fact that strong polymer-solvent attraction
can compress the coil from within \cite{Budkov1,Muzdalo,Kremer}. It should 
be noted that solvent mediated interactions between monomers that are due 
to the critical solvent density fluctuations only enhance this effect.

\begin{figure}[h]
\center{\includegraphics[width=1\linewidth]{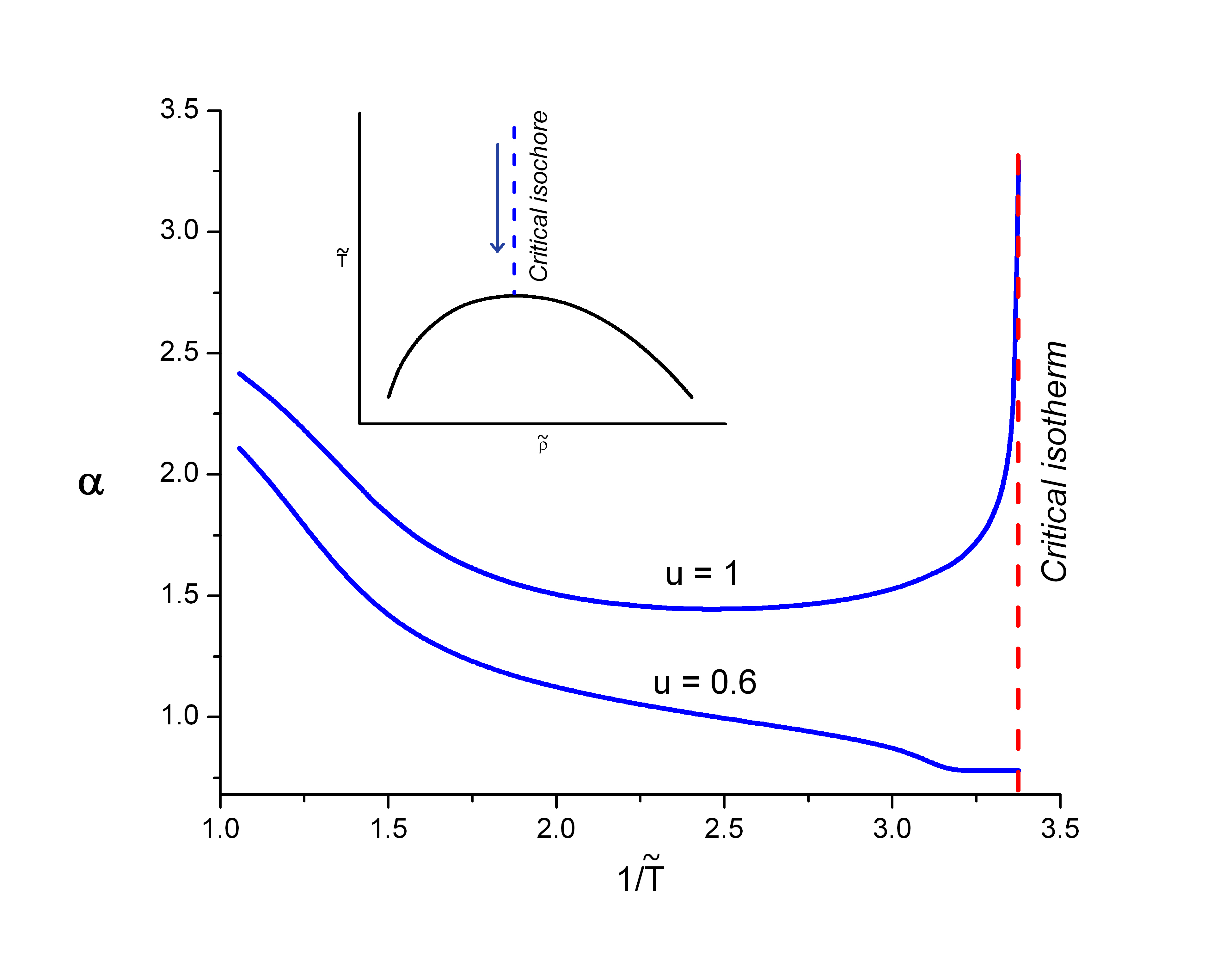}}
\caption{\sl The expansion factor $\alpha$ as a functions of reciprocal temperature $1/\tilde{T}$ for
different parameters $u=\epsilon_{ms}/\epsilon_{s}$ of polymer-solvent attraction.
On approaching the liquid-gas critical point of the solvent two qualitaively
different regimes of the expansion factor $\alpha$ take place. At weak polymer-solvent attraction
the polymer chain collapses at the critical point. When  $u$ exceeds a certain threshold
value the polymer chain undergoes at critical point a dramatic expansion.
Values are shown for $t=1$, $s=1$, $N=10^3$.}
\label{ris:3}
\end{figure}

\begin{figure}[h]
\center{\includegraphics[width=1\linewidth]{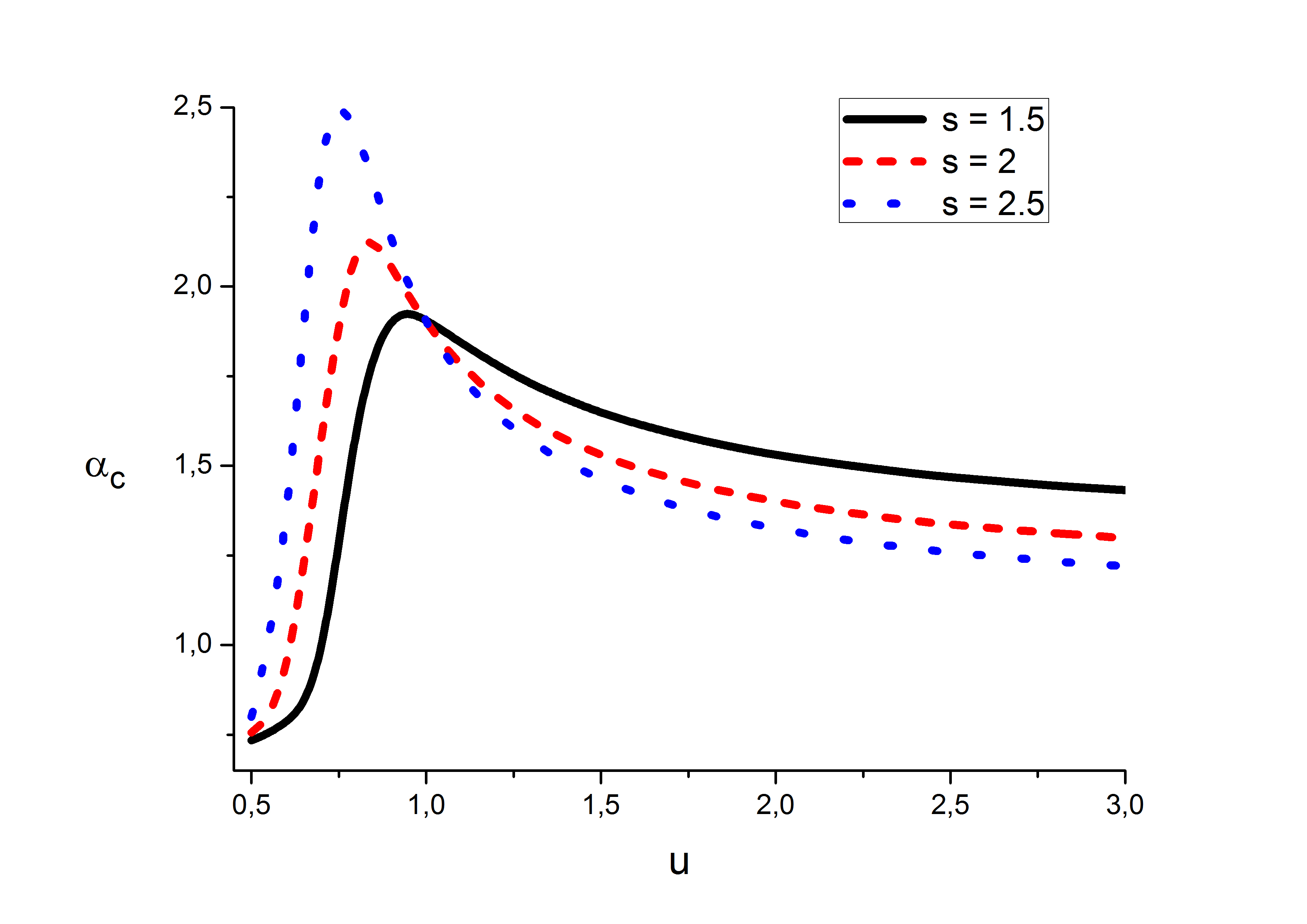}}
\caption{\sl  The expansion factor $\alpha_{c}=\alpha(\tilde{\rho}_{c},\tilde{T}_{c})$ at the liquid-gas critical point
as a function of polymer-solvent attraction  $u=\epsilon_{ms}/\epsilon_{s}$ shown for different values
of $s=\sigma_{m}/\sigma_{s}$. When the polymer-solvent attraction is weak the polymer is in a globular state.
%The increasing $u$ leads to a dramatic expansion of the polymer chain to a coiled state. 
For higher values of $u$  the polymer chain attains a coiled state. Increasing $u$ further leads to a shrinked  polymer coil again.
%the  Further increasing of 
%parameter $u$ leads to the shrinking of the polymer coil. 
Values are shown for $t=1$, $N=10^3$}
\label{ris:3}
\end{figure}
It is instructive to regard the conformational behavior of the polymer chain
for strong polymer-solvent attraction in a region above the critical point.
Particulary relevant is the case when the temperature is increased along an
isobar since this is an often realised situation in experiments.

As shown in Fig. 3, the expansion factor with increasing temperature
$T$ first monotonically decreases, attaining a local minimum,
and then monotonically increases. It should be emphasized, that the observed minimum
of expansion factor is most pronounced at the critical isobar, where the
fluctuations of solvent density are largest. Such behaviour of the expansion 
factor is in qualitatively agreement with results of Monte Carlo computer 
simulations reported in reference \cite{MC}.

\begin{figure}[h]
\center{\includegraphics[width=1\linewidth]{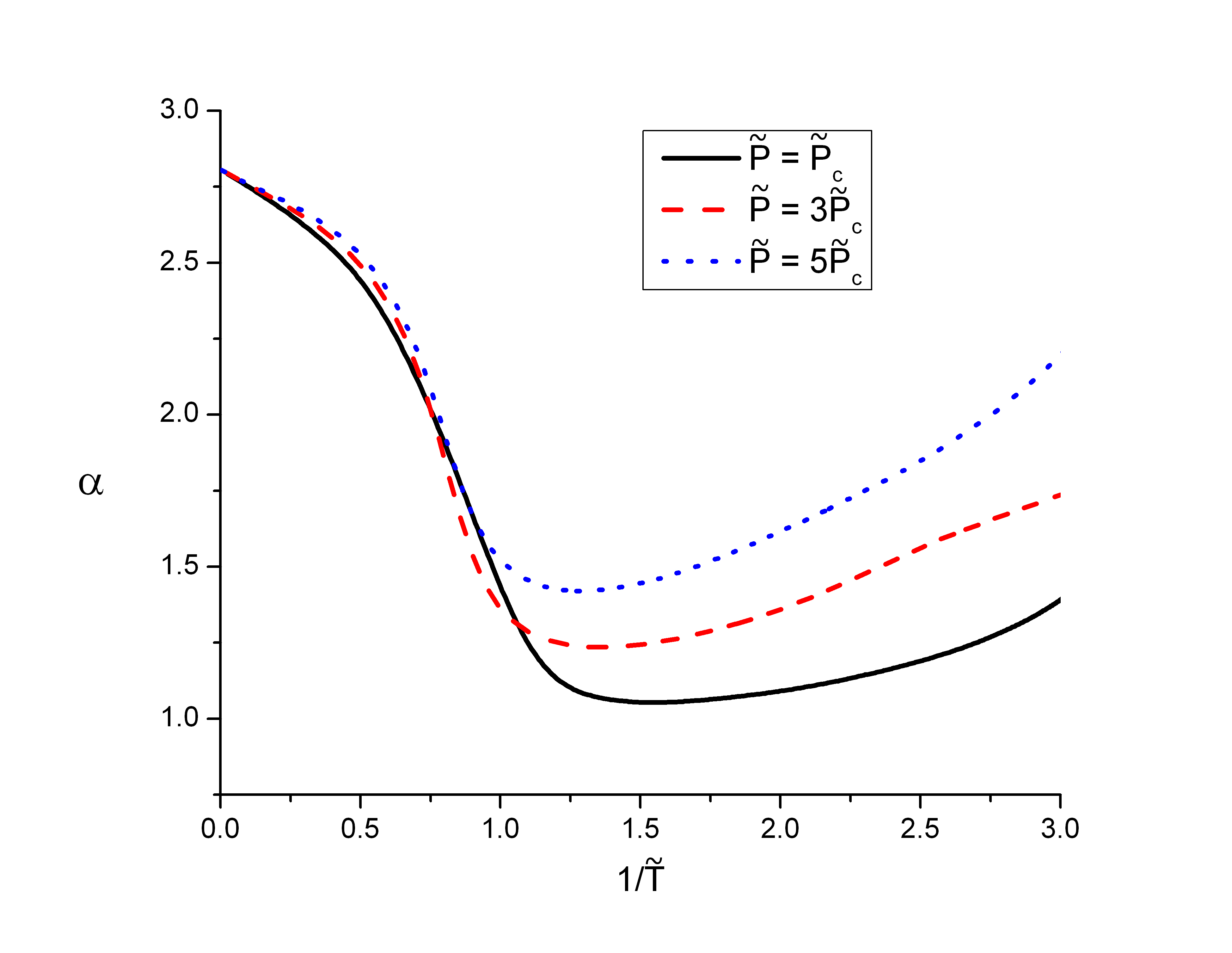}}
\caption{\sl  The expansion factor $\alpha$ as a function of reciprocal temperature $1/\tilde{T}$
shown for several isobars $\tilde{P}=\tilde{P}_{c},~2\tilde{P}_{c},~5\tilde{P}_{c}$ ($\tilde{P}_{c}=1/27$).
Values are shown for $t=1$, $N=10^3$, $u=2$, $s=1$.}
\label{ris:3}
\end{figure}

\section{Conclusion}
Employing a previously published \cite{Budkov1,Budkov2} self-consisted field theory,
incorporating a local solvent density renormalisation near the polymer chain and the effect
of solvent mediated fluctuation interaction monomer-monomer we have described the conformational
changes of a single polymer chain in low molecular weight solvent at critical
and supercritical parameters of state. Two qualitatively different regimes
were found depending on the attraction strength between the polymer and the solvent.
When the polymer-solvent attraction is weak progressing along
the critical isochore the polymer undergoes a collapse to a globular state at the liquid-gas
critical point. In contrast when the polymer-solvent attraction is strong the polymer first collapses,
attaing a minimum at higher than the critical temperature and strongly expands to a coiled conformation
at the critical temperature. These effects are caused by the critical fluctuations of the solvent density
in the vicinity of the liquid-gas critical point.

Previously reported results on conformational changes of a polymer chain have shown
two qualitative different regimes of the polymer chain -- collapse of the polymer
has been obtained by theory and simulations \cite{Erukhimovich,Sanchez,MC},
whereas expansion of the polymer has been obtained theoretically by several authors
\cite{Dua,Sekin}. The presented theory incorporating critical density fluctuations
and local solvent density renormalisation shows that both regimes are indeed possible,
depending on the monomer-solvent interaction parameter. We would like to stress that
similar results were obtained by Vasilevskaya et.al. in reference \cite{Khalatur}
by means of hybrid self-consistent MC/RISM method.

We would also like to mention that a more rigorous theory for the description of the coil-globule transitions 
has been developed in works of Lifshitz and co-authors \cite{Lifshitz1} based on the idea 
that the globule can be treated as a fragment of a semi-dilute polymer solution. 
In contrast to Flory type theories the behavior of a globule within the Lifshitz theory has been described 
in terms of the density functional theory, which allows to describe the globule's surface layer with implicit account 
of the solvent. Thus, Lifshitz theory introduces an additional length scale which can be interpreted as the thickness 
of the surface layer of the globule $R_s$ (an effective width of globule's $"$fringe$"$ \cite{Lifshitz1}) 
which smaller than $R_{g}$. It is clear that within simple Flory-type approaches the globule can undergo 
the dramatic expansion as whole in the case when $\xi\simeq const \times\tau^{-\nu}\geq R_{g}$ 
($\tau=(T-T_{c})/T_{c}$, $\nu$ is a one of the critical exponents). However within the 
more sophisticated Lifshitz theory one can investigate the case when the correlation length of the solution 
$\xi \simeq R_{s}$. Thus in this case one can expect the new effect when the expansion of 
globule's surface will take place. In other words, in this case the expansion of the globule will 
be at much wider temperature range than that is predicted within simple mean-field theories, 
namely at $\tau \leq   const\times  R_{s}^{-1/\nu}$. However, such speculations require more 
detailed investigations within the Lifshitz type approaches which can be a subject of forthcoming 
publications.

The described phenomena may be relevant for technologies where supercritical carbon dioxide
is used as a solvent to process or synthesize polymers \cite{Debenedetti,Kazaryan,Kikic,
Solubility_1,Solubility_2,Solubility_3}, produce polymer particles of controled dimensions \cite{Tomasko}
or to encapsulate bioactive molecules such as drugs, enzymes,
proteins into polymer particles \cite{Schmaljohann,Concheiro}. Controling
thermodynamic parameters of the solvent allows to change the polymer solubility, 
its conformation and in turn the concentration of solvent within the gyration volume.

\acknowledgments
The research leading to these results has received funding from the European
Union's Seventh Framework Program (FP7/2007-2013) under grant
agreement N//247500 with //project acronym "Biosol". The part concerning development
of theoretical model has been supported by Russian Scientific Foundation (grant N 14-33-00017).

\end{document}